\documentclass[aps,prd,preprintnumbers,nofootinbib,11pt]{revtex4} 

\usepackage{graphicx}
\usepackage{amsmath,amssymb,bm}

\newcommand{\calP}{{\cal P}}
\newcommand{\calR}{{\cal R}}

\begin{document}
\title{Inflation and Birth of Cosmological Perturbations
\footnote{Prepared for the Proceedings of {\it Relativity and Gravitation:
100 years after Einstein in Prague}, Prague, 25-29 June, 2012.}}

\author{Misao Sasaki}
\email{misao_AT_yukawa.kyoto-u.ac.jp}

\address{Yukawa Institute for Theoretical Physics, Kyoto University,
Kyoto 606-8502, Japan}

\preprint{YTIP-12-86} 

\begin{abstract}
We review recent developments in the theory
of inflation and cosmological perturbations produced from inflation.
After a brief introduction of the standard, single-field slow-roll inflation,
and the curvature and tensor perturbations produced from it,
we discuss possible sources of nonlinear, non-Gaussian perturbations
in other models of inflation.
Then we describe the so-called $\delta N$ formalism,
which is a powerful tool for
 evaluating nonlinear curvature perturbations on super Hubble scales.
\end{abstract}
\maketitle 

\section{Introduction}

One of the most successful applications of the theory of general 
relativity is cosmology. Over the past half century the big-bang theory
 of the universe, that the universe was born in an extremely hot
and dense state, expanded explosively and cooled down to the present
state, was observationally tested from various aspects and
it is now firmly established. According to the big-bang theory,
our universe is about 14 Giga years old, and
the universe was radiation-dominated in the beginning.
It became matter-dominated when the universe was about 100,000 years
old, which happens to be about the same time when the photons 
decoupled from baryons, and started to travel
freely until today, which are observed as the cosmic microwave
background (CMB) radiation. The epoch when the CMB photons were scattered 
last before they reach us forms a 3-dimensional hypersurface, and 
it is called the last scattering surface (LSS).

Despite its tremendous success,
there are still a couple of very basic problems that 
the big-bang theory cannot explain. One of them is the horizon problem
or perhaps better to be called the causality problem,
and the other the flatness problem or the entropy problem.

\subsection{Horizon problem}

Let us first consider the horizon problem.
The big-bang theory assumes an homogeneous and isotropic universe
on large scales. So the metric is assumed to be in the form,
\begin{eqnarray}
ds^2=-dt^2+a^2(t)d\sigma_{(3)}^2\,,
\label{Fmetric}
\end{eqnarray}
where $d\sigma_{(3)}^2$ is the 3-metric of
a constant curvature space with $K$ being the curvature,
${}^{(3)}R^{ij}{}_{km}=K(\delta^i_k\delta^j_m-\delta^i_m\delta^j_k)$.
A coordinate system that spans $d\sigma_{(3)}^2$ is said to be
comoving because an observer staying at a fixed point on the 3-space
is comoving with the expansion of the universe.
In this spacetime, the time-time component of the Einstein
equations, the Friedmann equation, is
\begin{eqnarray}
H^2=\frac{\rho}{3M_{pl}^2}-\frac{K}{a^2}\,; 
\quad H\equiv\frac{\dot a}{a}\,,
\label{Feq}
\end{eqnarray}
where $M_{pl}^2=(8\pi G)^{-1}$ in the units $\hbar=c=1$,
and the trace of the space-space components of the Einstein equations gives
\begin{eqnarray}
\frac{\ddot a}{a}=-\frac{\rho+3P}{3M_{pl}^2}\,,
\label{Eeq}
\end{eqnarray}
where $\rho$ is the energy density and $P$ is the pressure
of the universe. This latter equation shows that 
the expansion of the universe is always decelerating as
long as $\rho+3P>0$, which holds for both radiation $P=\rho/3$
and matter $P=0$. For simplicity, if we assume
a simple equation of state $P/\rho=w=$constant and $K=0$
(which should be a good approximation in the early universe
when $w=1/3$ since $\rho\propto a^{-3(1+w)}=a^{-4}$), one finds
\begin{eqnarray}
a\propto t^n\,;
\quad n=\frac{2}{3(1+w)}<1
\quad\mbox{for}~w>-\frac{1}{3}\,.
\end{eqnarray}
This result may be regarded as a consequence of the attractive
nature of the gravitational force.

\begin{center}
\begin{figure}[htb]
  \begin{center}
  \includegraphics[width=9cm]{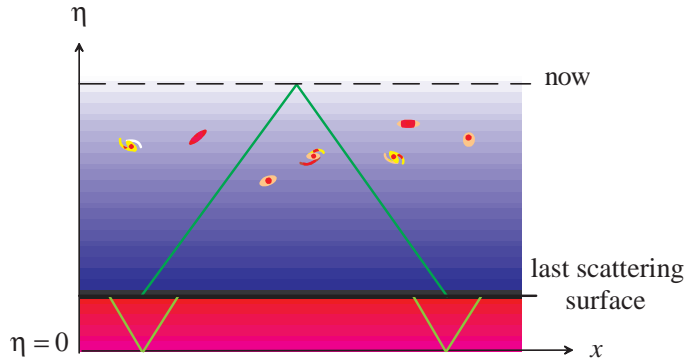}
  \end{center}
  \caption{\label{fig:hprob}
Horizon problem. The conformal time
of the last scattering surface $\eta_{LSS}$ from $\eta=0$
is about 1/30 of that of today $\eta_0$.}
\end{figure}
\end{center}

Now we introduce the conformal time $d\eta=dt/a(t)$,
and rewrite the metric as
\begin{eqnarray}
ds^2=a^2(\eta)d\hat s^2\,;
\quad d\hat s^2=-d\eta^2+d\sigma_{(3)}^2\,.
\end{eqnarray}
Since the conformal transformation of the metric
does not change the causal structure, the static metric 
$d\hat s^2$ perfectly describes the causal
structure of the universe. If the range of $\eta$ were infinite
to the past, there would be no horizon problem.
The problem is that the conformal time is finite in the past
if $w>-1/3$ or $\rho+3P>0$, because
\begin{eqnarray}
\eta=\int_{0}^t\frac{dt'}{a(t')}\propto\int_0^t\frac{dt'}{t'{}^n}\,;
\quad n=\frac{2}{3(1+w)}\,.
\end{eqnarray}
This implies that the size of lightcone emanating from
a point at the beginning of the universe when $\eta=0$
will cover only a finite fraction of spacetime.
Since the comoving distance traveled by light is 
equal to the corresponding conformal time interval,
the comoving radius of the causally connected region on the 
LSS is equal to its conformal time $\eta_{LSS}$.
From the fact that the LSS is located at redshift $z\sim10^3$
and the universe is approximately matter-dominated since then,
one finds that this region will cover only a tiny fraction 
(about $10^{-3}$ sr) of the sky.
This is the horizon problem (see Fig.~\ref{fig:hprob}).

The solution is clear: The horizon problem disappears
if the conformal time is either infinite in the past or the beginning
of the universe $\eta=0$ is extended sufficiently back in time
to cover the whole visible universe.
Since the comoving radius of the visible universe on the LSS
is $\eta_0-\eta_{LSS}$ where $\eta_0$ is the conformal
time today, the problem is solved if $\eta_{LSS}>\eta_0-\eta_{LSS}$.
In Einstein gravity, this means that
the equation of state must be $w<-1/3$ or the expansion of
the universe must be accelerating ($\ddot a>0$) for a sufficient
lapse of time in the very early universe.

Here we should note that solving the horizon problem does {\it not\/}
mean explaining the homogeneity and isotropy of the universe. 
As clear from the above argument, we had to assume 
the homogeneity and isotropy of the universe to pose the horizon
problem. This point is very often misunderstood in the literature.

\subsection{Flatness problem}

Again we assume a spatially homogeneous and isotropic universe,
Eq.~(\ref{Fmetric}). The Friedmann equation~(\ref{Feq}) tells us
that the curvature term $K/a^2$ is completely negligible in the
early universe when $\rho\propto a^{-4}$. 
Conversely, if the curvature term was of the same order of
magnitude as the density at an epoch in the early universe,
the universe must have either collapsed (if $K>0$)
or become completely empty (if $K<0$) by now.

Alternatively, since the energy density is dominated by radiation
in the early universe and so is the entropy of the universe,
the problem may be rephrased as the existence of huge
entropy within the curvature radius of the universe,
\begin{eqnarray}
S=T^3\left(\frac{a}{\sqrt{|K|}}\right)^3
\approx T_0^3\left(\frac{a_0}{\sqrt{|K|}}\right)^3
>T_0^3H_0^3\approx 10^{87}\,,
\end{eqnarray}
where $T_0\approx2.7$\,K is the CMB temperature today~\cite{Mather:1993ij}
and $H_0\approx 72$\,km/s/Mpc is the Hubble constant~\cite{Freedman:2010xv}.
Hence the flatness problem may be called the
entropy problem.

It is then apparent that the solution to the flatness
problem needs huge entropy production at a sufficiently
early stage of the universe.

\subsection{Inflation as a solution to horizon and flatness
problems}

A simple and perhaps the best solution to the horizon and flatness
problems is given by the inflationary
universe~\cite{Sato:1980yn,Guth:1980zm}.
 Let us assume that the universe was dominated by
a spatially homogeneous scalar field. For a minimally coupled
canonical scalar field $\phi$, we have
\begin{eqnarray}
\rho=\frac{1}{2}\dot\phi^2+V(\phi)\,,
\quad
P=\frac{1}{2}\dot\phi^2-V(\phi)\,,
\end{eqnarray}
so $\rho+3P=2\bigl(\dot\phi^2-V(\phi)\bigr)$.
Hence if $\dot\phi^2<V(\phi)$, we may have 
accelerated expansion. In particular, if 
the energy density is dominated by the potential
energy, $\dot\phi^2\ll V(\phi)$, the motion of
the scalar field can be ignored within a few expansion
times $\sim H^{-1}$, and the universe expands almost
exponentially,
\begin{eqnarray}
H^2\approx\frac{\rho}{3M_{pl}^2}\approx\mbox{constant}.
\end{eqnarray}
The curvature term $K/a^2$ becomes completely negligible.

Thus if the universe is dominated by the potential energy,
or the {\it vacuum energy\/}, and the potential energy is 
converted to radiation after a sufficient lapse of time of
such a stage, a huge entropy is produced and the horizon 
and flatness problems are solved simultaneously.

\section{Slow-roll inflation and vacuum fluctuations}
\label{sec:slowroll}

There have been a number of proposals for inflationary models.
Among others, a simplest class of models, and which
explains the observational data almost perfectly, is 
the slow-roll infation~\cite{Linde:1981mu,Albrecht:1982wi,Linde:1983gd}. 
The field equation for $\phi$ and the Friedmann equation are
\begin{eqnarray}
\ddot\phi+3H\dot\phi+V'(\phi)=0\,,
\quad
H^2=\frac{1}{3M_{pl}^2}\left[\frac{1}{2}\dot\phi^2+V(\phi)\right]\,,
\end{eqnarray}
where we have justifiably neglected the curvature term.

\begin{center}
\begin{figure}[htb]
  \begin{center}
  \includegraphics[width=10cm]{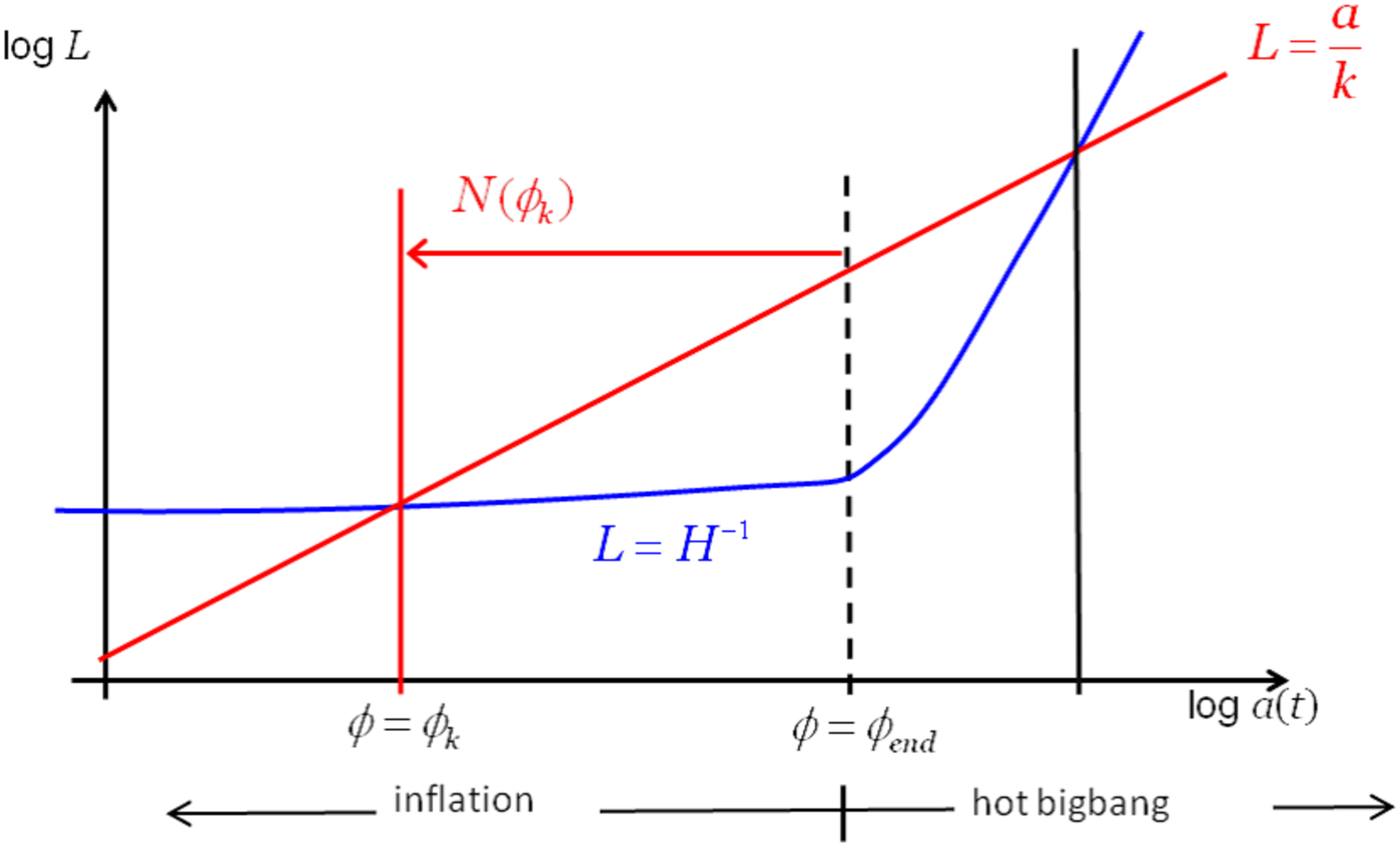}
  \end{center}
  \caption{\label{fig:scale} The Hubble radius$L=H^{-1}$ 
and the length scale $L=a/k$ of a comoving wavenumber $k$ 
in the inflationary cosmology,
 and the definition of the number of $e$-folds $N(\phi)$.
}
\end{figure}
\end{center}

The standard slow-roll condition consists of two assumptions.
One is that $\ddot\phi$ is negligible compared to $3H\dot\phi$
in the field equation, that is, the equation of motion is
friction-dominated. The other is that the kinetic term
$\dot\phi^2/2$ is negligible compared to the potential term
$V$ in the energy density. Under this condition we have
\begin{eqnarray}
\dot\phi=-\frac{V'(\phi)}{3H}\,;
\quad
H^2=\frac{V}{3M_{pl}^2}\,.
\label{sreq}
\end{eqnarray}
Then the potential energy dominance implies
\begin{eqnarray}
\epsilon\equiv-\frac{\dot H}{H^2}=
\frac{\dfrac{3}{2}\dot\phi^2}{\dfrac{1}{2}\dot\phi^2+V}
\approx\frac{3\dot\phi^2}{2V}\approx
\frac{M_{pl}^2}{2}\frac{V'{}^2}{V^2}\equiv\epsilon_V \ll1\,,
\label{epsilon}
\end{eqnarray}
that is, the universe is expanding almost exponentially,
and the friction-dominated equation of motion
$|\ddot\phi/(3H\dot\phi)|\ll1$ implies
\begin{eqnarray}
\frac{V''}{3H^2}\approx M_{pl}^2\frac{V''}{V}\equiv\eta_V\ll1\,.
\label{etaV}
\end{eqnarray}
The single-field slow-roll inflation satisfies these conditions.

The important property of slow-roll inflation is that Eq.~(\ref{sreq})
is completely integrable since $H$ is a function of $\phi$.
In particular, there is one-to-one correspondence between $\phi$ and $t$.
So instead of the cosmic time $t$ we may measure the time
in terms of the value of the scalar field.

Here we introduce a quantity which plays a very important
role in the dynamics of slow-roll inflation, namely the number
of $e$-folds counted {\it backward\/} in time, say from the end
 of inflation to an epoch during inflation,
\begin{eqnarray}
\frac{a(t_{end})}{a(t)}=\exp[N(t\to t_{end})]
~\to~N=N(\phi)=\int_{t(\phi)}^{t_{end}}Hdt\,.
\label{Ndef}
\end{eqnarray}
Its important property is that by definition it does not depend
on how and when the inflation began. As shown in Fig.~\ref{fig:scale},
$N$ is uniquely determined in terms of the value of the scalar field
(up to a constant which depends on the choice of an epoch from
which $N$ is computed), and one can associate $N$ with the time
at which a given comoving wavenumber $k$ crossed the Hubble radius,
$k=aH$, at which the value of the scalar field was $\phi_k$; $N=N(\phi_k)$.
As we shall see below, this turns out to be an essential quantity
for the evaluation of the curvature perturbation from inflation.

\subsection{Curvature perturbation}
Let us now consider the curvature perturbation produced from inflation.
It arises from the quantum vacuum fluctuations of the inflaton field $\phi$.
Since a rigorous derivation would take too much space, here we give
an intuitive, rather hand-waving derivation. We caution
that it could well lead to an incorrect result if used blindly.

The vacuum fluctuations of the inflaton field with a comoving wave number
$k$ is given simply by its positive frequency function, $\varphi_k$.
Because of the condition $V''/H^2\ll 1$, on scales $k/a\gg H$, 
the inflaton field fluctuation behaves like a minimally coupled
massless scalar. Hence we have
\begin{eqnarray}
\left|\langle\delta\phi|\vec k\rangle\right|^2=|\varphi_k|^2\,,
\quad
\varphi_k\sim \frac{1}{a^{3/2}\sqrt{2\omega_k}}e^{-i\omega_kt}\,;
\quad \omega_k=\frac{k}{a}\gg H\,.
\label{vacfluc}
\end{eqnarray}
As the universe expands the physical wavenumber decreases exponentially
and becomes smaller than the Hubble parameter, $k/a<H$, or the
physical wavelength exceed the Hubble radius.
Then the oscillations of $\varphi_k$ are frozen.
This could be regarded as ``classicalization' of the
quantum fluctuations. Not that this is merely an interpretation.
In a more rigorous sense, freezing of the mode function
is a process toward infinite squeezing of the vacuum state.

Setting $a=k/H$ in Eq.~(\ref{vacfluc}) gives
\begin{eqnarray}
\varphi_k\sim\frac{H}{\sqrt{2k^3}}\,;\quad \frac{k}{a}\ll H\,.
\end{eqnarray}
Therefore the mean square amplitude in unit logarithmic interval of 
$k$ is
\begin{eqnarray}
\langle\delta\phi^2\rangle_k\equiv
\frac{4\pi k^3}{(2\pi)^3}|\varphi_k|^2\approx 
\left(\frac{H}{2\pi}\right)_{k/a=H}^2\,.
\end{eqnarray}
Inclusion of the non-trivial evolution of the background
spacetime and the coupling of the scalar field fluctuation
with the metric fluctuation do not change the above estimate
if we interpret $\delta\phi$ in the above as those
evaluated on the flat slicing, that is, on hypersurfaces
on which the spatial scalar curvature remains unperturbed.

It is known that the curvature perturbation on the comoving 
hypersurface $\calR_c$ is conserved if the perturbation is 
adiabatic~\cite{Kodama:1985bj}.
The comoving hypersurface is defined as a surface of uniform $\phi$.
 Then the gauge transformation from the flat slicing to the comoving 
slicing gives the relation between $\calR_c$ and $\delta\phi$,
\begin{eqnarray}
\calR_c=-\frac{H}{\dot\phi}\delta\phi\,.
\label{calRdphi}
\end{eqnarray}
Since this is conserved for $k/a<H$,
the spectrum of the comoving curvature perturbation in
unit logarithmic interval of $k$ is given by
\begin{eqnarray}
\calP_\calR(k)\equiv\frac{4\pi k^3}{(2\pi)^3}|\calR_k|^2
\approx\left(\frac{H^2}{2\pi\dot\phi}\right)_{k/a=H}^2\,.
\label{calRspec}
\end{eqnarray}
A rigorous, first-principle derivation of the above result was
first done in \cite{Mukhanov:1985rz,Sasaki:1986hm}.

The important relation of the above result with
the number of $e$-folds was first pointed out
in~\cite{Starobinsky:1986fxa}: If we rewrite Eq.~(\ref{Ndef}) as
\begin{eqnarray}
N=\int_t^{t_{end}}Hdt
=\int_\phi^{\phi_{end}}\frac{H}{\dot\phi}d\phi\,,
\end{eqnarray}
we find
\begin{eqnarray}
\delta N(\phi_k)
=\left[\frac{\partial N}{\partial\phi}\delta\phi\right]_{k/a=H}
=\left[-\frac{H}{\dot\phi}\delta\phi\right]_{k/a=H}=\calR_c\,,
\end{eqnarray}
provided that we identify $\delta\phi$ with
the scalar field fluctuation evaluated on the flat hypersurface.
This is called the $\delta N$ formula.

The $\delta N$ formula implies that we only need the knowledge of
the background evolution to obtain the power spectrum of the
comoving curvature perturbation, once we know the amplitude of
the quantum fluctuations of the scalar field at the horizon crossing
(i.e. when $k/a=H$). It is quite generally 
given by $H/(2\pi)$ in slow-roll inflation. With careful
geometrical considerations, the $\delta N$ formula can be 
extended to general multi-field inflation~\cite{Sasaki:1995aw},
\begin{eqnarray}
\calP_\calR(k)=\left(\frac{H}{2\pi}\right)^2||\nabla N||^2
;\quad ||\nabla N||^2\equiv 
G^{ab}(\phi)\frac{\partial N}{\partial\phi^a}
\frac{\partial N}{\partial\phi^b}\,,
\label{mdeltaN}
\end{eqnarray}
where $G^{ab}$ is the field space metric and it is assumed that
the vacuum expectation values are given by
\begin{eqnarray}
\langle\delta\phi^a\delta\phi^b\rangle=G^{ab}\left(\frac{H}{2\pi}\right)^2\,.
\end{eqnarray}
The nonlinear generalization of the $\delta N$
formalism will be discussed in Sec.~\ref{sec:deltaN}.

\subsection{Tensor perturbation}

There are not only vacuum fluctuations of the inflaton field
but also those of the transverse-traceless part of the metric,
$\partial^ih_{ij}^{TT}=\delta^{ij}h_{ij}^{TT}=0$,
that is, the tensor perturbation or gravitational wave degrees of freedom.
If we construct the second-order action for $h_{ij}^{TT}$,
we find
\begin{eqnarray}
S\sim \frac{M_{pl}^2}{8}\int d^4x\sqrt{-g}(\dot h_{ij}^{TT})^2+\cdots.
\end{eqnarray}
To quantize $h_{ij}^{TT}$ it is convenient to normalize
the kinetic term to the canonical form. This gives
\begin{eqnarray}
S\sim \frac{1}{2}\int d^4x\sqrt{-g}(\dot\phi_{ij})^2+\cdots\,;
\quad \phi_{ij}\equiv \frac{M_{pl}}{2}h_{ij}^{TT}.
\end{eqnarray}
If one writes down the field equation for $\phi_{ij}$,
one finds its mode function $\phi_k$ obeys exactly the same equation
as the one for a minimally coupled massless scalar field,
\begin{eqnarray}
\ddot\phi_k+3H\dot\phi_k+\frac{k^2}{a^2}\phi_k=0\,.
\end{eqnarray}
Since there are two independent degrees of freedom in $\phi_{ij}$,
the power spectrum of the tensor perturbation $h_{ij}^{TT}$ is obtained as
\begin{eqnarray}
\calP_T(k)=\frac{4}{M_{pl}}\times 2\times
\frac{4\pi k^3}{(2\pi)^3}|\phi_k|^2=\frac{8H^2}{(2\pi)^2M_{pl}^2}\,.
\label{tensorspec}
\end{eqnarray}

Taking the ratio of the tensor spectrum to the curvature perturbation
spectrum, we find~\cite{Sasaki:1995aw}
\begin{eqnarray}
r\equiv\frac{\calP_T}{\calP_\calR}\leq 8|n_T|=-2\frac{\dot H}{H^2}\,,
\label{tsratio}
\end{eqnarray}
where $n_T$ is the tensor spectral index,
$n_T=d\ln\calP_T(k)/d\ln k$, and the equality holds for
the case of single-field slow-roll inflation.
This is a consistency relation in general slow-roll inflation.
 As a proto-type example, if we consider chaotic 
inflation~\cite{Linde:1983gd}, we expect to have $r\sim0.1$.

The important point to be kept in mind is that
the existence of the vacuum fluctuations of the 
tensor part of the metic is a proof of the existence of quantum gravity.
These fluctuations exist in any theory of gravity that respects
general covariance, apart from possible inessential modifications of the
 spectrum. Thus a clear detection of the tensor spectrum will be
a confirmation of not only the inflationary universe but also 
quantum gravity. 

\section{Origin of non-Gaussianity} 

The standard, single-field slow-roll inflation predicts that
the curvature perturbation is a Gaussian random field
and it has an almost scale-invariant spectrum.
This seems to fit the current observational data quite 
well~\cite{Komatsu:2010fb},
it is quite possible that the actual model turns out to
be non-standard. Maybe it is multi-field, maybe non-slow-roll
and/or non-canonical. In such a case, the curvature perturbation
may become non-Gaussian. Search for possible non-Gaussian signatures
in the primordial curvature perturbation has become
one of the important directions in observation in recent 
years~\cite{Komatsu:2009kd}.

Here we consider possible origins of non-Gaussianity in the
curvature perturbation. Essentially one can classify the
origins into three categories: (1) Self-interactions
of the inflaton field and/or non-trivial vacua,
(2) multi-field dynamics, and
(3) nonlinearity in gravity.

The non-Gaussianities of the first category are
generated on subhorizon scales during inflation,
hence they are of quantum field theoretical origin.
Those of the second category are usually generated
on superhorizon scales either during or after
inflation, and they are due to nonlinear coupling
of the scalar field to gravity. Since they are
generated on superhorizon scales, they are of
classical origin. Finally those of the third category
are due to nonlinear dynamics in general relativity.
Hence they are generated after the scale of interest
re-enters the Hubble horizon.
Since the last category is not really primordial in nature, 
let us focus on the first two categories.

\subsection{Non-Gaussianity from self-interaction/non-trivial vacuum}

It is known that conventional self-interactions by the potential 
are ineffective~\cite{Maldacena:2002vr}. 
This can be seen by considering chaotic inflation, for example.
In the simplest case of a quadratic potential, $V=m^2\phi^2/2$,
 the inflaton is actually a free field apart from the
interaction through gravitation perturbations.
But the gravitational interaction is Planck-suppressed,
i.e., it is always suppressed by a factor $O(M_{pl}^{-2})$.
In the case of a quartic potential, $V=\lambda\phi^4$,
it is known that $\lambda$ should be extremely small
$\lambda\sim10^{-15}$ in order for it to be consistent
with observation.

Thus some kind of unconventional self-interaction
is necessary. A popular example is the case
of a scalar field with a non-canonical kinetic term
such as DBI inflation~\cite{Alishahiha:2004eh}.
In this case the kinetic term takes the form,
\begin{eqnarray}
K\sim f^{-1}(\phi)\sqrt{1-f(\phi)\dot\phi^2}\equiv f^{-1}\gamma^{-1}\,.
\end{eqnarray}
If we expand this perturbatively, 
\begin{eqnarray}
K=K_0+\delta_1K+\delta_2K+\delta_3K+\cdots,
\end{eqnarray}
we will find 
\begin{eqnarray}
\delta_2K\propto \gamma^3\,,
\quad
\delta_3K\propto \gamma^{3+2}\,,
\end{eqnarray}
since $\delta\gamma=\gamma^3\delta X$ where
$X\equiv f\dot\phi^2/2$. If we regard the third order
part as the interaction, the above implies
that the scalar field fluctuation will be expressed
qualitatively as
\begin{eqnarray}
\delta\phi\sim\delta\phi_0+\gamma^2\delta\phi_0^2+\cdots,
\end{eqnarray}
where $\delta\phi_0$ is the free, Gaussian fluctuation.
Thus the non-Gaussianity in $\delta\phi$ may become
large if $\gamma$, which mimics the Lorentz factor, 
is large~\cite{Mizuno:2009cv}.

A non-trivial vacuum state is another source of non-Gaussianity.
If the universe were a pure de Sitter spacetime,
gravitational interaction would be totally negligible
in vacuum, except for the effect due to graviton (tensor mode)
loops. This may be regarded as due to the maximally symmetric
nature of the de Sitter space, $SO(4,1)$,
which has the same number of degrees of symmetry as the
 Poincare (Minkowski) symmetry.
In slow-roll inflation, the de Sitter symmetry is slightly broken.
Nevertheless the effect induced by this symmetry breaking is
small because it is suppressed by the slow-roll 
parameter $\epsilon=-\dot H/H^2$.

However, if the vacuum state does not respect the de Sitter
symmetry, there can be a large non-Gaussianity.
Such a deviation from the quasi-de Sitter vacuum, usually
called the Bunch-Davies vacuum, may occur in various
situations, studied e.g. in \cite{Chen:2008wn,Flauger:2009ab}.

\subsection{Non-Gaussianity from multi-field dynamics}

Non-Gaussianity may appear if the energy momentum tensor
depends nonlinearly on the scalar field even if the fluctuation
of the scalar filed itself is Gaussian. This effect is generally
important when the fluctuations are on superhorizon scales,
i.e., the characteristic wavelength is larger than the Hubble radius.
It is small in single-field slow-roll models because the linear
approximation is valid to high accuracy~\cite{Salopek:1990jq}, 
generically suppressed by
the slow-roll parameter $\eta_V$ defined in Eq.~(\ref{etaV}).

\begin{center}
\begin{figure}[htb]
  \begin{center}
  \includegraphics[width=5cm]{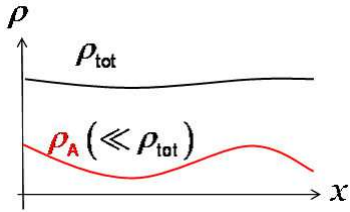}
  \end{center}
  \caption{\label{fig:muldensity} 
An illustration of the energy density configuration
in the multi-field case. The density of the $A$-matter/field
$\rho_A$ may vary nonlinearly without significantly affecting the
total energy density.
}
\end{figure}
\end{center}

For multi-field models, however, the contribution to the
energy momentum tensor from some of the fields can be 
highly nonlinear as depicted in Fig.~\ref{fig:muldensity}. 
The important property of non-Gaussianity
in this case is that it is always of the spatially local type.
Namely, to second order in nonlinearity, the curvature
perturbation will take the form~\cite{Komatsu:2001rj},
\begin{eqnarray}
\calR_c(x)=\calR_{c,0}(x)+\frac{3}{5}f_{NL}^{\rm local}\,\calR_{c,0}^2(x)\,,
\label{fNLdef}
\end{eqnarray}
where $\calR_{c,0}$ is the Gaussian random field and $f_{NL}^{\rm local}$
is a constant representing the amplitude of non-Gaussianity.
The factor 3/5 in front of $f_{NL}^{\rm local}$ is due to a historical reason.
The reason why it is of local type is simply causality: No information
can propagate over a length scale greater than the Hubble horizon scale.

Observationally, this type of non-Gaussianity can be tested
by using the so-called squeezed type templates where
one of the wavenumbers, say $k_1$ in the bispectrum
$B({\vec k}_1,{\vec k}_2,{\vec k}_3)$ is much smaller than
the other two, $k_1\ll k_2\approx k_3$~\cite{Komatsu:2009kd},
and there are a few observational indications that $f_{NL}^{\rm local}$
is actually non-vanishing. For example, the WMAP 7 year data analysis
gave a one-sigma bound $11<f_{NL}^{\rm local}<53$
 ($68\%$ CL)~\cite{Komatsu:2010fb}.

\section{$\bm{\delta N}$ formalism}
\label{sec:deltaN}

As mentioned in Sec.~\ref{sec:slowroll}, the $\delta N$ formalism is
a powerful tool to evaluate the comoving curvature perturbation
on superhorizon scales. It then turned out that it can be easily
extended to the evaluation of nonlinear, non-Gaussian curvature
 perturbations~\cite{Lyth:2004gb,Lyth:2005fi}.
Let us recapitulate its definition and properties:
\begin{list}{}{}
\item[(1)]
$\delta N$ is the perturbation in the number of $e$-folds counted
{\it backward\/} in time from a fixed final time, say $t=t_f$,
to some initial time $t=t_i$.
\item[(2)]
The final time $t_f$ should be chosen such that the evolution of the
universe has become unique by that time, i.e., the universe has
reached the adiabatic limit. Then the hypersurface $t=t_f$ should
be identified with a comoving (or uniform density) slice,
and the initial hypersurface $t=t_i$ should be identified
with a flat slice.
\item[(3)]
$\delta N$ is equal to the conserved (nonlinear) comoving curvature
perturbation on superhorizon scales at $t>t_f$.
\item[(4)]
By definition, it is nonlocal in time. However, because of its
purely geometrical definition, it is valid independent of which 
theory of gravity one considers, provided that the adiabatic limit
is reached by $t=t_f$.
\end{list}

\begin{center}
\begin{figure}[htb]
  \begin{center}
  \includegraphics[width=11cm]{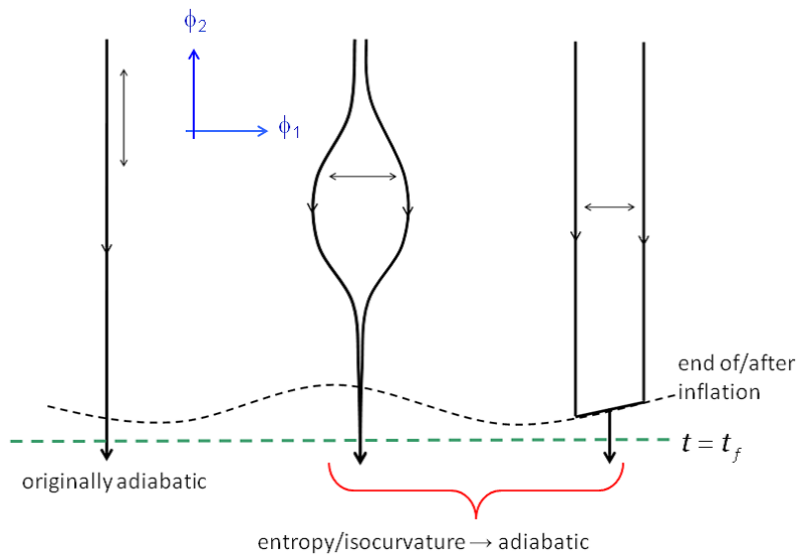}
  \end{center}
  \caption{\label{fig:deltaN} 
Three different types of $\delta N$.
The field space $(\phi_1,\phi_2)$ in the figure represents the degrees of 
freedom in the initial condition of the universe.
The adiabatic limit is defined to be the stage by which
all the trajectories converge to a unique one.}
\end{figure}
\end{center}

There are various kinds of sources that generate $\delta N$.
They may be classified into three types, as depicted in Fig.~\ref{fig:deltaN}.
The left one describes a perturbation along the evolutionary
trajectory of the universe.
This case is the same as that of single-field slow-roll inflation, 
in which the comoving curvature perturbation is conserved all the way
until it re-enters the horizon.
The middle one is the case when a small difference in the initial data
develops into a substantial difference in $\delta N$. Typically this is
realized when there is some instability orthogonal to the trajectory,
like the case when the scalar field moves along a ridge.
This type of sources of $\delta N$ usually induces a feature in the 
spectrum and/or bispectrum of the curvature perturbation.
The right one represents the case when the perturbation orthogonal to
the trajectory does not contribute to the curvature perturbation
until or after the end of inflation, but $\delta N$ is generated due to
a sudden transition that brings the universe into an adiabatic stage.
Typical examples are curvaton models~\cite{Lyth:2001nq,Moroi:2001ct,Sasaki:2006kq}
and multi-brid inflation models~\cite{Sasaki:2008uc,Naruko:2008sq}.

\begin{center}
\begin{figure}[htb]
  \begin{center}
  \includegraphics[width=11cm]{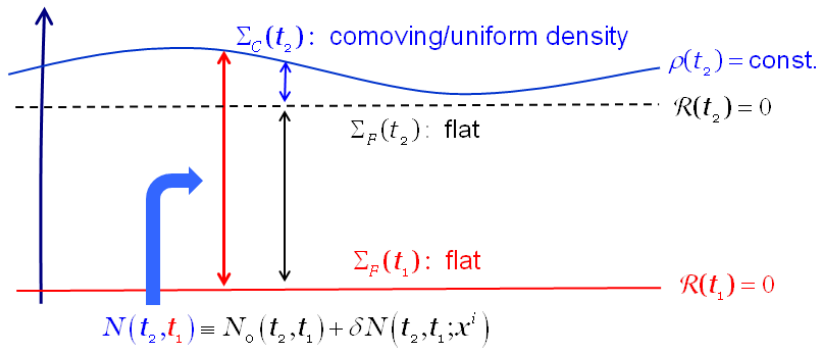}
  \end{center}
  \caption{\label{fig:NLdeltaN} 
Definition of nonlinear $\delta N$. It is defined as
the perturbation in the number of $e$-folds from an
initial flat slice to a final comoving slice.}
\end{figure}
\end{center}

Here, for the sake of completeness, let us present the precise definition
of the nonlinear $\delta N$ formula. See Fig.~\ref{fig:NLdeltaN}.
It is based on the leading order
approximation in the spatial gradient expansion or the separate universe 
approach~\cite{Lyth:2004gb}, where spatial derivatives are assumed
to be negligible in comparison with time derivatives.
At leading order of the spatial gradient expansion, if
we express the spatial volume element as 
$\sqrt{{}^{(3)}\gamma}=a^3(t)\exp[3\calR(t,x)]$ where $a(t)$ is 
the scale factor of a fiducial homogeneous and isotropic universe,
we easily find that the perturbation in the number of $e$-folds
along a comoving trajectory between two hypersurfaces 
$t=t_1$ and $t=t_2$ is given by
\begin{eqnarray}
\delta N(t_2,t_1;x^i)=\calR(t_2,x^i)-\calR(t_1,x^i)\,,
\label{deltaNgeneral}
\end{eqnarray}
where $x^i$ are the comoving coordinates. Here we note that
this is purely a geometrical relation. It has nothing to do with 
any equations of motion.

First we fix the final hypersurface $t=t_2$. It should be taken at
the stage when the evolution of the universe has become unique.
That is, there exists no isocurvature perturbation any longer that could
develop into an adiabatic perturbation at later epochs. 
Thus the comoving curvature perturbation is conserved at $t>t_2$.
In the context of the concordance $\Lambda$CDM model of the universe,
this corresponds to the final radiation-dominated stage of the universe.

Next we choose the initial slice $t=t_1$. It should be chosen to be flat.
Here `flat' means that the perturbation in the spatial volume element
vanishes. Namely, the flat slice is defined as a hypersurface on which 
$\calR=0$.
We note that despite its name, the scalar curvature vanishes only in the
linear theory limit: It is non-vanishing in general in the nonlinear case.

Applying the above choice of the initial and final hypersurfaces
to Eq.~(\ref{deltaNgeneral}), it is trivial to see that we have
\begin{eqnarray}
\delta N(t_2,t_1;x^i)=\calR_c(t_2,x^i)\,.
\label{NLdNform}
\end{eqnarray}
Now by assumption $\calR_c$ is conserved at $t>t_2$. So it is the
quantity we want to evaluate. This completes the derivation of
the nonlinear $\delta N$ formula.

As mentioned above, since Eq.~(\ref{deltaNgeneral}) is a pure
geometrical relation, so is the nonlinear $\delta N$ formula~(\ref{NLdNform}).
This is the reason why it can be applied to any theory of gravity
as long as it is a geometrical (i.e. general covariant) theory.

Of course, the above definition tells us nothing about how to 
evaluate it in practice. In this respect, we have a very fortunate
situation in the case of inflationary cosmology. It is the fact that
the evaluation of the quantum fluctuations of the inflaton
field, whether it is single- or multi-component,
can be most easily done in a gauge in which the time slicing
is chosen to be flat~\cite{Sasaki:1995aw}.
Thus we can choose the initial slice to be an epoch when
the scale of our interest has just exited the horizon
during inflation. Let the fluctuations of a multi-component scalar
field on the flat slice at $t=t_1$ to be $\delta\phi^a$.
Then assuming that the values of the scalar field determine
the evolution of the universe completely, which is the case for
slow-roll inflation, the nonlinear $\delta N$ can be simply evaluated as
\begin{eqnarray}
\delta N=N(\phi^a+\delta\phi^a)-N(\phi^a)\,,
\label{dNphi}
\end{eqnarray}
where $N(\phi^a)$ is the $e$-folding number of the fiducial
background. In particular, to second order in $\delta\phi^a$,
we obtain
\begin{eqnarray}
\calR_c=\delta N=\frac{\partial N}{\partial\phi^a}\delta\phi^a
+\frac{1}{2}
\frac{\partial^2N}{\partial\phi^a\partial\phi^b}\delta\phi^a\delta\phi^b
+\cdots\,,
\label{dNphi2}
\end{eqnarray}
Comparing this with Eq.~(\ref{fNLdef}), we see that the
curvature perturbation takes a bit more complicated form that
the simplest form. Nevertheless if we consider the bispectrum,
i.e., the Fourier component of the three-point function
$\langle\calR_c(x_1)\calR_c(x_2)\calR_c(x_3)\rangle$, we find
there is a quantity that exactly corresponds to
$f_{NL}^{\rm local}$ defined in Eq.~(\ref{fNLdef}).
Namely~\cite{Lyth:2005fi},
\begin{eqnarray}
\frac{3}{5}f_{NL}^{\rm local}
=\frac{G^{ab}G^{cd}N_aN_{bc}N_d}{2\left(||\nabla N||^{2}\right)^2}\,;
\quad N_a\equiv\frac{\partial N}{\partial\phi^a}\,,
\ N_{ab}\equiv\frac{\partial^2N}{\partial\phi^a\partial\phi^b}\,.
\label{dNphiG}
\end{eqnarray}

Before concluding this section, we mention the fact that the $\delta N$
 formalism does not require the scalar field fluctuations to be Gaussian.
In fact, except for the last equation in the above, Eq.~(\ref{dNphiG})
which assumes the Gaussianity of $\delta\phi^a$, 
the general $\delta N$ formula~(\ref{dNphi}) or its second order 
version~(\ref{dNphi2}) can be used for non-Gaussian 
$\delta\phi^a$~\cite{Byrnes:2007tm}.
Such a case may happen, for example, in multi-field DBI inflation.

\section{Summary}

It has been about 30 years since the inflationary universe was 
first proposed, and there is increasing observational evidence 
that inflation did take place in the very early universe.
Among others, the measured CMB temperature anisotropy
is fully consistent with the predictions of inflation
that the primordial curvature perturbation spectrum is
almost scale-invariant and it is statistically Gaussian.

Inflation also predicts a scale-invariant tensor spectrum,
and if the energy scale of inflation is high enough as
in the case of chaotic inflation, the tensor-scalar ratio
$r$ can be as large as 0.1. If this is the case, the
tensor perturbation will be detected in the near future,
and it will confirm not only the inflationary universe
but also quantum gravity.

Even if the tensor perturbation will not be detected,
there may be other interesting signatures of inflation.
Non-Gaussianity from inflation is attracting attention
as one of those signatures that can distinguish
or constrain models of inflation significantly. 

We discussed that the origins of primordial non-Gaussianities
may be classified into three categories, according to
different length scales on which different mechanisms
are effective:
\begin{list}{}{}
\item[(1)]
Quantum theoretical origin on subhorizon scales during inflation.
\item[(2)]
Classical nonlinear scalar field dynamics on superhorizon scales
during or after inflation.
\item[(3)]
Nonlinear gravitational dynamics after the horizon re-entry.
\end{list}
In particular we argued that non-Gaussianities
in the second case are always of spatially local type.
We then mentioned that there are three different kinds of 
situations in which such local non-Gaussianities can be generated, 
and described in some detail a very efficient method to compute them, 
namely, the $\delta N$ formalism.

Apparently identifying properties of primordial
non-Gaussianities in the observational data is extremely 
important for understanding the physics of the early universe.
Here we mentioned only the bispectrum or the 3-point function.
But if it is detected, higher order $n$-point functions may
become important as a model discriminator. Other types
of non-Gaussianity discriminators may also become necessary.

What is important is that we are now beginning to test observationally
the physics of the very early universe,
the physics at an energy scale closer to the Planck 
scale, at a scale that can never be attained in high energy
accelerator experiments.

Cosmology has become not only a precision science,
but now it constitutes a truly indispensable part of
fundamental physics. 
General relativity is the backbone
of cosmology. I wonder what Einstein would say
if he were here in this very exciting era --
100 years after he visited Prague.

\section*{Acknowledgements}
I am very grateful to the organizers of the
conference, "Relativity and Gravitation, 100 years
after Einstein in Prague", particularly to Jiri Bicak,
who kindly invited me to this meeting, and who
accorded me a warm hospitality. 
I am also grateful to Laila Alabidi for careful reading
of the manuscript and very useful comments.
This work was supported in part by JSPS Grant-in-Aid for
 Scientific Research (A) No.~21244033, and by
Monbukagaku-sho Grant-in-Aid for the Global COE programs, 
``The Next Generation of Physics, Spun from Universality 
and Emergence" at Kyoto University.
{}


\section*{References}

\begin{thebibliography}{}

\bibitem{Mather:1993ij} 
  J.~C.~Mather {\it et al.},
  Astrophys.\ J.\  {\bf 420}, 439 (1994).

\bibitem{Freedman:2010xv} 
  W.~L.~Freedman and B.~F.~Madore,
  Ann.\ Rev.\ Astron.\ Astrophys.\  {\bf 48}, 673 (2010)
  [arXiv:1004.1856 [astro-ph.CO]].

\bibitem{Sato:1980yn} 
  K.~Sato,
  Mon.\ Not.\ Roy.\ Astron.\ Soc.\  {\bf 195}, 467 (1981).

\bibitem{Guth:1980zm} 
  A.~H.~Guth,
  Phys.\ Rev.\ D {\bf 23}, 347 (1981).

\bibitem{Linde:1981mu} 
  A.~D.~Linde,
  Phys.\ Lett.\ B {\bf 108}, 389 (1982).

\bibitem{Albrecht:1982wi} 
  A.~Albrecht and P.~J.~Steinhardt,
  Phys.\ Rev.\ Lett.\  {\bf 48}, 1220 (1982).

\bibitem{Linde:1983gd} 
  A.~D.~Linde,
  Phys.\ Lett.\ B {\bf 129}, 177 (1983).

\bibitem{Kodama:1985bj} 
  H.~Kodama and M.~Sasaki,
  Prog.\ Theor.\ Phys.\ Suppl.\  {\bf 78}, 1 (1984).

\bibitem{Mukhanov:1985rz} 
  V.~F.~Mukhanov,
  JETP Lett.\  {\bf 41}, 493 (1985).

\bibitem{Sasaki:1986hm} 
  M.~Sasaki,
  Prog.\ Theor.\ Phys.\  {\bf 76}, 1036 (1986).

\bibitem{Starobinsky:1986fxa} 
  A.~A.~Starobinsky,
  JETP Lett.\  {\bf 42}, 152 (1985).


\bibitem{Sasaki:1995aw} 
  M.~Sasaki and E.~D.~Stewart,
  Prog.\ Theor.\ Phys.\  {\bf 95}, 71 (1996).
  [astro-ph/9507001].


\bibitem{Komatsu:2010fb} 
  E.~Komatsu {\it et al.}  [WMAP Collaboration],
  Astrophys.\ J.\ Suppl.\  {\bf 192}, 18 (2011)
  [arXiv:1001.4538 [astro-ph.CO]].

\bibitem{Komatsu:2009kd} 
  E.~Komatsu {\it et al.},
  arXiv:0902.4759 [astro-ph.CO].
\\
See also articles in the focus section,
(ed.) M.~Sasaki and D.~Wands,
``Non-linear and non-Gaussian cosmological perturbations,''
  Class.\ Quant.\ Grav.\  {\bf 27}, 120301 (2010).

\bibitem{Maldacena:2002vr} 
  J.~M.~Maldacena,
  JHEP {\bf 0305}, 013 (2003)
  [astro-ph/0210603].

\bibitem{Alishahiha:2004eh} 
  M.~Alishahiha, E.~Silverstein and D.~Tong,
  Phys.\ Rev.\ D {\bf 70}, 123505 (2004)
  [hep-th/0404084].

\bibitem{Mizuno:2009cv} 
  S.~Mizuno, F.~Arroja, K.~Koyama and T.~Tanaka,
  Phys.\ Rev.\ D {\bf 80}, 023530 (2009)
  [arXiv:0905.4557 [hep-th]].

\bibitem{Chen:2008wn} 
  X.~Chen, R.~Easther and E.~A.~Lim,
  JCAP {\bf 0804}, 010 (2008)
  [arXiv:0801.3295 [astro-ph]].
\bibitem{Flauger:2009ab} 
  R.~Flauger, L.~McAllister, E.~Pajer, A.~Westphal and G.~Xu,
  JCAP {\bf 1006}, 009 (2010)
  [arXiv:0907.2916 [hep-th]].

\bibitem{Salopek:1990jq} 
  D.~S.~Salopek and J.~R.~Bond,
  Phys.\ Rev.\ D {\bf 42}, 3936 (1990).

\bibitem{Komatsu:2001rj} 
  E.~Komatsu and D.~N.~Spergel,
  Phys.\ Rev.\ D {\bf 63}, 063002 (2001)
  [astro-ph/0005036].


\bibitem{Lyth:2004gb} 
  D.~H.~Lyth, K.~A.~Malik and M.~Sasaki,
  JCAP {\bf 0505}, 004 (2005).
  [astro-ph/0411220].

\bibitem{Lyth:2005fi} 
  D.~H.~Lyth and Y.~Rodriguez,
  Phys.\ Rev.\ Lett.\  {\bf 95}, 121302 (2005)
  [astro-ph/0504045].

\bibitem{Lyth:2001nq} 
  D.~H.~Lyth and D.~Wands,
  Phys.\ Lett.\ B {\bf 524}, 5 (2002)
  [hep-ph/0110002].

\bibitem{Moroi:2001ct} 
  T.~Moroi and T.~Takahashi,
  Phys.\ Lett.\ B {\bf 522}, 215 (2001)
  [Erratum-ibid.\ B {\bf 539}, 303 (2002)]
  [hep-ph/0110096].

\bibitem{Sasaki:2006kq} 
  M.~Sasaki, J.~Valiviita and D.~Wands,
  Phys.\ Rev.\ D {\bf 74}, 103003 (2006)
  [astro-ph/0607627].

\bibitem{Sasaki:2008uc} 
  M.~Sasaki,
  Prog.\ Theor.\ Phys.\  {\bf 120}, 159 (2008)
  [arXiv:0805.0974 [astro-ph]].
\bibitem{Naruko:2008sq} 
  A.~Naruko and M.~Sasaki,
  Prog.\ Theor.\ Phys.\  {\bf 121}, 193 (2009)
  [arXiv:0807.0180 [astro-ph]].

\bibitem{Byrnes:2007tm} 
  C.~T.~Byrnes, K.~Koyama, M.~Sasaki and D.~Wands,
  JCAP {\bf 0711}, 027 (2007)
  [arXiv:0705.4096 [hep-th]].

\end{thebibliography}
\end{document}